\documentclass[12pt]{iopart}

\newcommand{\gws}{gravitational waves~}
\newcommand{\gw}{gravitational wave~}
\usepackage{iopams}
\usepackage{graphicx}
\usepackage[usenames]{color}
\usepackage{amssymb}

\begin{document}

\title[Progress towards Gravitational Wave Astronomy]{Progress towards Gravitational Wave Astronomy}

\author{Maria Alessandra Papa}

\address{Max Planck Institute for Gravitational Physics\\
Am Muehlenberg 1\\
14476 Golm\\
Germany
}
\begin{abstract}
I will review the most recent and interesting results from gravitational wave detection experiments, concentrating on recent results from the LIGO Scientific Collaboration (LSC). I will outline the methodologies utilized in the searches, explain what can be said in the case of a null result, what quantities may be constrained. I will compare these results with  prior expectations and discuss their significance. As I go along I will outline the prospects for future improvements.
\end{abstract}

\pacs{01.30.Cc,04.80.Nn,95.85.Sz,07.05.kf}


\section{Introduction}

\begin{figure}[!htbp]
\begin{center}
\includegraphics[width=0.85\textwidth]{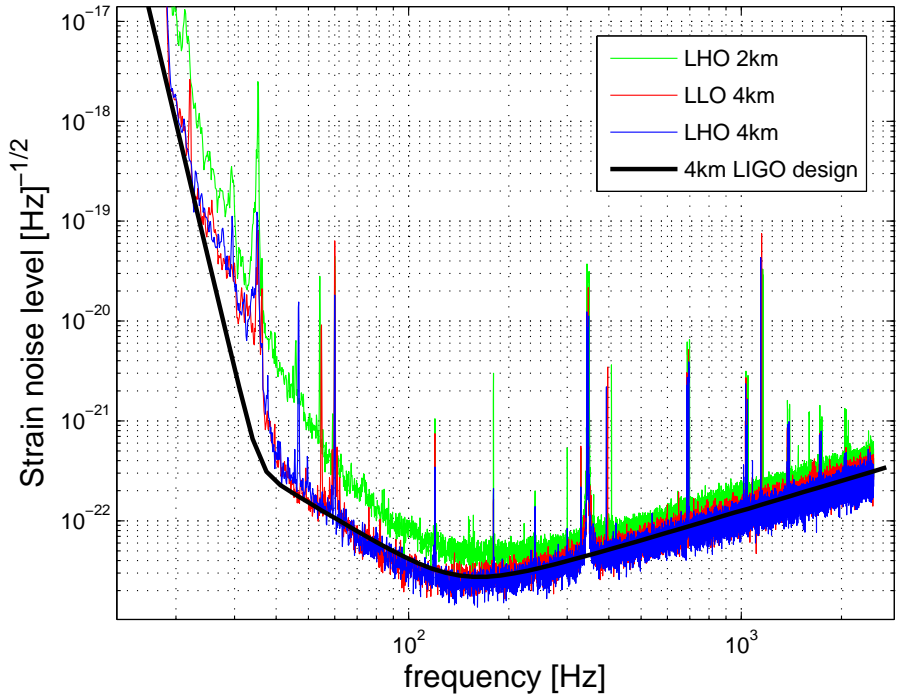}
\caption{LIGO sensitivity during its 5th science run. The black solid curve shows the design sensitivity goal.}
\label{fig:ligosens}
\end{center}
\end{figure}

It is a very exciting time for gravitational wave data analysts: LIGO has just completed its 5th science run (S5), lasting 2 years and having achieved and surpassed its design sensitivity goal \cite{whitcomb,lscExp07}. Fig. \ref{fig:ligosens} shows typical noise spectral density curves during the S5 run. The solid curve shows the design sensitivity goal for the longest baseline detectors. 

A shorter baseline detector, GEO600 \cite{geo600}, has been pooling its data with LIGO since the first science run in 2002 and the two projects have integrated their data analysis activities within the LSC. The experimental efforts of GEO and of LIGO have a long history of cooperation, and GEO600 also serves as a testbed for the technologies that will be used in the next generations of detectors. The GEO600 detector is located near Hannover, Germany, and after S5 it is being operated in ``astrowatch-mode'' to ensure some coverage for loud gravitational wave events, while the network of long baseline detectors are off the air for upgrades.

During the last 5 months of S5 the Virgo \cite{Virgo} detector has also joined forces with the LSC, with a data sharing agreement that will carry through to the next science run of the enhanced detectors, expected to be operational in 2009, with a sensitivity improvement of a factor of $\approx 2$ with respect to the original design sensitivity goal.

Data analysis for interferometric \gw detectors builds on the legacy that many years of analysis of resonant bar detector data have contributed to this field. The results detailed in \cite{IGEC2} use over 130 days of joint observation by the three detectors AURIGA, NAUTILUS and EXPLORER setting a duty factor milestone that remains unbeaten to date.

The sensitivity band of LIGO extends between 50 Hz and 1500 Hz. In this band we expect gravitational wave signals from compact binary systems, at various stages of their evolution: during their inspiral, the coalescence/merger phase and from the oscillations of the object that forms after the merger. We expect \gws to be emitted in association with supernova collapse events; we also expect emission of continuous gravitational waves and a stochastic gravitational wave background. Given the expected rarity and low signal-to-noise-ratios (SNR), the searches for the different signals can be carried out largely independently of one another. 


Searches for inspiral signals utilize matched filtering, which is the optimal detection method for linear systems and gaussian noise under the assumption that the form of the signal that one looks for is well known in advance. Similar techniques could in principle be adopted in searching for continuous waves, if the problem were not severely computationally limited. Short lived signals, occuring during catastrophic events, are not as well modeled, and the use of matched filtering methods is not possible. A stochastic background is not a deterministic signal that one can predict as a function of time. Cross correlation techniques are utilized to detect this type of signal.

In the next sections I will review a sample of interesting results for each type of search. I will outline the basic search method and, in the case of null results, discuss what quantities may be constrained by the available data, and what the significance of the constraint is.  

\section{Inspiral searches}\label{cbc}

Binary systems of compact objects evolve in orbits that gradually shrink in time, due to the emission of gravitational radiation \cite{hulsetaylor1,hulsetaylor2}.  As the orbits shrink the frequency of the \gw signal increases, faster and faster, producing a chirp-like signal. The exact time-frequency evolution of the signal depends on a number of parameters, but the large timescales are set by the total mass of the system. The amplitude of the signal increases with the mass of the system and as the orbits shrinks. 

Systems with masses up to 200 solar masses are expected to emit signals with significant energy content in LIGO's band. However the waveform can only be confidently predicted in the adiabatic regime where a post-Newtonian expansion may be used to model the evolution of the phase and amplitude of the signal. This regime happens at different frequencies for different mass systems. 

The inner-most stable circular orbit for a non-spinning system of $2.8$ solar masses is estimated to be at 1600 Hz and scales  in a manner inversely proportional to the mass. 
It is only for non-spinning systems with total mass smaller than $3$ solar masses that the waveforms in the LIGO band can be accurately predicted by post-Newtonian models  for the phase and the amplitude of the \gw signal. However, outside of this range of parameters template waveforms can still be constructed which capture the features of a large class of signals and enable detection. This is an area under active investigation (\cite{S3spinning,s3s4,interpreting} and references therein) and for the purpose of this review I will not discuss the problem of accurate waveform modelling. 

Inspiral searches are carried out by analyzing separately the data from each detector via a matched filtering technique, by setting a threshold on the resulting detection statistic and by selecting candidates exceeding the threshold which are ``not too close'' to each other in parameter space. Such candidate lists from the different detectors are then compared and coincident events are identified. These events are kept only if they pass signal-based vetos, and they constitute the surviving candidates of the analysis. 

Signal-based vetos are statistics which are constructed to answer the question ``does this event, that has a high detection statistic significance, really look like a putative signal ?''. This is a question that the detection statistic does not fully answer. In fact, the same high detection statistic may indicate both that a few very high data samples match a few points of the waveform or that many samples match the waveform across the entirety of its duration. Obviously the latter is more likely to be the signal that one is searching for.

Historically, signal-based vetos have been devised for searches for short lived signals, whose templates excite many disturbances and noise transient in the data. The traditional signal-based veto used in inspiral searches is the $\chi^2$ veto \cite{chi2}. This veto tests the hypothesis that the energy of the event is distributed in frequency in a manner that is consistent with the excited template waveform. Other types of signal-based vetos have been developed, for example based on the shape of the $\chi^2$ statistic as a function of time.

The same search procedure is applied to data streams whose samples have been artificially time-shifted. This produces off-source candidate lists which are used to estimate the background of events. The on-source distribution is compared to the background in order to assess the presence of significant deviations between the two and thus the presence of a detection. In any case, the most significant (highest SNR) candidates are followed-up by in-depth case-by-case scrutiny. This stage aims at challenging one's confidence level regarding the candidate: it exercises one's ability to understand the details of how the instrument was functioning at any given time. It also makes use of ancillary signal-based information to judge the consistency in the appearance of the putative waveform in the different data streams.

Fig. 3 of \cite{s3s4} shows the distributions of on-source events overlayed on the estimated backgrounds for one of the most recently released inspiral search results . In the absence of a detection upper limits are set on the rate of events \cite{cbcMethod1,interpreting}. In particular Fig. 6 of \cite{s3s4} shows the 90\% confidence upper limits on the rate of binary inspiral events from systems having total mass between $\approx$ 1 and 80 solar masses, from S4 data. The upper limits lie in the range $R=$ 0.5-4.9 events per year per $L_{10}$ across the parameter space, where $L_{10}$ is $10^{10}$ times the solar blue light luminosity. These event rates are a few orders of magnitude higher that the predictions for the expected rates for such events.

It is straightforward to translate a number of events-per-$L_{10}$-per-year quantity $R$ into the number of events that we expect to detect with a given search:
\begin{equation}
{\rm{expected~ \# ~of~ events}}~ = ~ R ~ \times T ~ \times C({\rm{search}}) 
\label{eq:expectedNumevents}
\end{equation}
where $T$ is the length of the data used in the search and $C$ is  a measure of the effective number of galaxies that the search can reach, expressed in $L_{10}$.

\begin{figure}[!htbp]
\begin{center}
\includegraphics[width=0.8\textwidth]{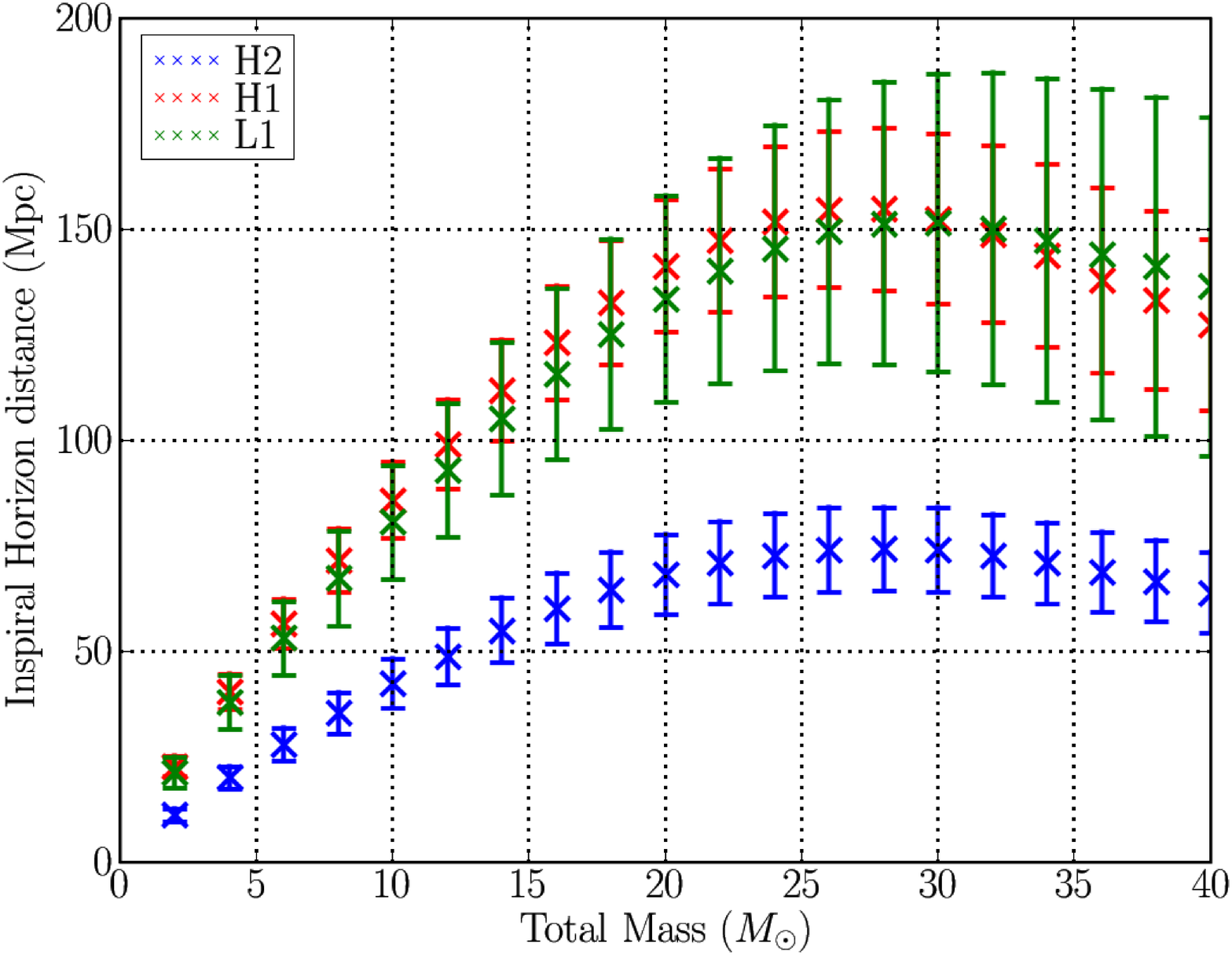}
\caption{Typical horizon distance during the S5 run as a function of the total mass of the binary system. {\it Courtesy of the LSC}.}
\label{fig:horizonS5}
\end{center}
\end{figure}

The sensitivity of a search may be characterized by its horizon distance $d_H$. This is the distance at which an optimally located and oriented equal mass binary system is expected to produce a signal with matched-filter SNR = 8. For real systems the reach of a search in general will not exceed its horizon distance. Fig.\ref{fig:horizonS5} shows estimates of the horizon distance during the S5 run.

A catalogue of galaxies has been developed to provide the total luminosity $C(d)$ contained within a sphere of radius $d$ from the Earth  \cite{catalogue}. However this is not quite what one needs in Eq. \ref{eq:expectedNumevents}. In fact not all the systems, i.e. the luminosity, within the horizon distance sphere are equally GW-visible because of their different positions and orientations with respect to the detectors. For this reason in \cite{catalogue} the cumulative luminosity has also been derived as a function of the horizon distance, $C(d_H)$. This function is shown in Fig.[7] of \cite{catalogue}. The value of $d_H$ to use in $C(d_H)$ is given by the appropriate plot of the type shown in Fig. \ref{fig:horizonS5}. 

The sensitivity of the S5 searches is such that one should not expect to surely detect a binary inspiral signal with about a year of data. In fact
Eq. \ref{eq:expectedNumevents} yields for the S5 expected detection rates 1 event per 400 to 25 years for 1.4-1.4 solar mass systems; 1 event every 2700 to 20 years for 5-5 solar mass systems and 1 event every 450 to 3 years for 10-10 solar mass systems. Enhanced detectors are expected to achieve an improvement in strain sensitivity of a factor of $\approx$ 2. With a horizon distance of 60 Mpc to neutron star systems the expected rates grow to 1 event every 60 to 4 years of actual observing time. Advanced detectors operating at a horizon distance of 450 Mpc to neutron star systems, bring the expected detection rates between several to order hundred events per year of observing time.

\section{Burst searches}\label{bursts}

There are many circumstances in which short bursts of \gws are expected, lasting from a few ms to a few seconds involving the merger phase of a binary system or the collapse of a stellar core. Due to the nature of this type of event, typically catastrophic, the shape of the \gw signal is poorly known. Matched filtering techniques cannot be utilized and one resorts to the use of more robust, but generally less sensitive, excess-power and power-tracking based methods. 

Typically these excess-power and power-tracking techniques are applied separately to each data stream and a detection statistic is computed by combining the data from the different detectors. Candidates are then identified based on having exceeded a significance threshold, having passed a number of signal-based vetos, as well as vetos based on environmental and auxiliary channel information. These veto procedures significantly reduce the false alarm rate. As for the inspiral searches, the same pipeline is applied to time-shifted data in order to estimate the background of accidentals. The most recently released result analyzes the S4 data, finds no surviving candidates and sets a 90\% upper limit on the rate of detectable events of $\sim$ 1 per week \cite{burstS4}.

The sensitivity of the burst search pipelines is measured by injecting putative fake signals in the data and then by measuring the efficiency with which one recovers them. Unlike the case of signals from the adiabatic phase of a binary inspiral, for burst signals there exists no analytical waveform and the wave shapes come from numerical studies. 

\begin{figure}[!htbp]
\begin{center}
\includegraphics[width=0.95\textwidth]{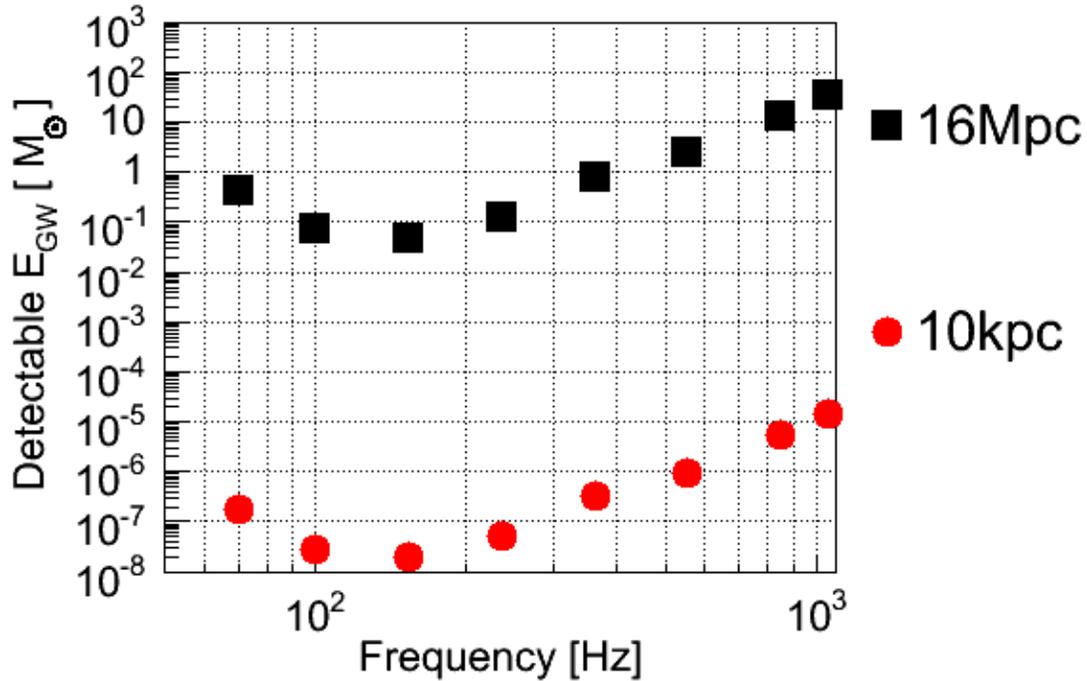}
\caption{Energy in emitted in gravitational waves that would generate a sine-gaussian signal detectable with 50\% efficiency, as a function of the frequency of the sine-gaussian waveform.{\it Courtesy of the LSC.}}
\label{fig:burstEff}
\end{center}
\end{figure}

Empirically it has been shown that the most relevant features of most numerical waveforms are well captured by a suitable gaussian or sine-gaussian waveform. Each waveform at the detector can be parametrized by an amplitude 
\begin{equation}
h_{rss}=\sqrt {\int {h_+ (t)^2 + h_\times (t)^2 }}~dt.
\label{eq:hrss}
\end{equation} 
By assuming isotropic emission, for sine-gaussians with a quality factor $\gg 1$, one can derive an expression that connects the emitted gravitational wave energy $E_{GW}$ with  $h_{rss}$, the frequency of the signal $f_0$ and the distance to the source $d$: 
\begin{equation}
E_{GW}={d^2 c^3 \over 4G}(2\pi f_0)^2 h_{rss}^2
\label{eq:Egw}
\end{equation} 
From detection efficiency studies one can derive the probability of  detection as a function of $h_{rss}$ for various sine-gaussian signals. Through Eq. \ref{eq:Egw} one can then draw $E_{GW}(f_0 | d)$ curves: the energy emitted in a high quality factor sine-gaussian waveform at a frequency $f_0$, by a source at a distance $d$,  at a level detectable with a certain confidence. Preliminary estimates of the reach of S5 burst searches are shown in Fig.\ref{fig:burstEff}: close to the best sensitivity spot of the instruments, a 50\% detection efficiency is achieved for signals generated by converting of order 5\% of a solar mass at the distance of the Virgo cluster, or $\sim 2\times 10^{-8}$ of a solar mass at the Galactic center. Estimates of the expected amplitude of burst signals, mostly from numerical simulations, vary quite widely and scenarios exist which predict emission that is detectable in S5. For example \cite{baker06} predicts for black hole mergers the emission of up to 3\% of solar masses in gravitational waves. A system of this type formed by two 50 solar mass black holes at $\approx$ 100 Mpc would produce \gws which could be detected with 50\% efficiency in S5.

\section{Triggered searches}

An interesting category of searches are the ones triggered by electromagnetic observations, for example X-ray and gamma-ray bursts. Within this category the most interesting result is certainly that on the implications for the origin of GRB 070201 from LIGO observations \cite{grb070201}.

GRB 070201 was an intense, hard GRB localized within an area which included one of the spiral arms of the M31 Galaxy. It is commonly accepted that short GRBs may be produced in the merger phase of binary neutron star systems (BNS) or neutron star-black hole binaries (NSBH). During S5 the reach of a search for a 1.4-1.4 solar mass inspiral in S5 was around 30 Mpc, see Fig. \ref{fig:horizonS5}. Since M31 is at approximately 800 kpc, this GRB could well have been associated with a detectable \gw signal.  At the time of the GRB the Hanford detectors were taking data. An inspiral search was carried out on that data for systems with component masses in the range 1-3 and 1-40 solar masses respectively.  A search for an unmodeled burst was also carried out, by crosscorrelating the data stream from the two detectors. No signal was found in either search. The inspiral search excluded the possibility that the GRB was due to a binary neutron star or NSBH inspiral signal in M31 with very high confidence (greater than 99\%). It also excluded various companion mass - distance ranges significantly further than M31, as shown in Fig. 3 of \cite{grb070201}. The unmodeled burst search produced 90\% upper limits on $h_{rss}$ at different frequencies, which could then be re-cast as upper limits on the isotropic \gw emission at the distance of M31. The most stringent upper limit is $\approx 8\times 10^{50}$ erg, which is orders of magnitude larger that the estimated energy release in gamma rays at the same distance. A soft gamma ray repeater (SGR) flare event in M31 is consistent with the gamma-ray energy release and is not ruled out by the \gw analysis \cite{ofek}.

\section{Searches for a stochastic gravitational wave background}

An isotropic stochastic background of gravitational radiation is expected due to the superposition of many unresolved signals, both of cosmological and astrophysical origin. The background is described by a function $\Omega_{GW}(f)$, which is proportional to the energy density in \gws per logarithmic frequency interval. For most models this quantity may be modeled by a power law. Most commonly a flat spectrum is assumed. Searches are performed by appropriately combining the cross-correlations of two detectors (for a detailed illustration see \cite{stochLSCMethod}). The resulting crosscorrelation value is compared with the background distribution for the same quantity, obtained by the time-shifting method described in the context of binary inspiral searches. In the case of a null result an upper limit is placed on $\Omega_{GW}(f)$, and most commonly for flat spectrum on the coefficient $\Omega_0$. 

The optimal filter involves multiplying the Fourier transform of the data by the overlap reduction function, which is a function that depends on the relative orientation and distance of the two detectors \cite{stochMethod}.  The overlap reduction function models the decrease in the sensitivity of the search for distant and/or misaligned detectors with respect to the co-aligned and co-located ones. In principle the most sensitive stochastic background searches are performed with coaligned, nearby instruments, such as the 2km and 4km detectors at Hanford. The gain with respect to using the Hanford-Livingston pair is of order of 10 in $\Omega_{GW}$ above 50 Hz. However, since many components of the two Hanford detectors share the same physical facilities, their data hosts a variety of correlated environmental noise and disturbances. This has de facto precluded the use of the two co-aligned co-located Hanford instruments. Investigations are underway to alleviate this loss by estimating the narrow band noise correlations in the two co-located instruments \cite{fotopoulos}. 

Above 200 Hz the reduction in the envelope of the overlap reduction function depends on proximity of the instruments and favours the Virgo-GEO600 pair. When factoring in the sensitivity of the instruments above 200 Hz a transatlantic network of detectors operating at their design sensitivities is about a factor of $\sqrt 2$ more sensitive than the LSC-only network \cite{stochNetworks}. 

The most recent results from searches for isotropic backgrounds come from the analysis of the S4 LIGO data and for a flat \gw spectrum put a 90\% Bayesian upper limit at $\Omega_{GW} \times \left[ {H_0\over {72 {\rm {km s^{-1} Mpc^{-1}}}}}\right] < 6.5\times 10^{-5}$, in the frequency range 50-150 Hz \cite{S4stoch}. This limit is still above the one that may be inferred from measurements of light-element abundances, WMAP data and the big bang nucleosynthesis model, but it is expected that the data from the S5 run will probe values of $\Omega_{GW}$ below this. 

A method for searching for a non-isotropic, broad-band background has recently been developed \cite{radiometerMethod}  and applied to S4 data \cite{S4radiometer}. The detection statistic is a directional crosscorrelation, which may be re-expressed in terms of the coefficient $H_\beta$ which characterizes the searched \gw background power-law spectrum $H(f)=H_\beta \left( {f\over 100 {\rm{Hz}}} \right)^\beta$. Results on the S4 data show no evidence of a detection and constrain a white \gw spectrum between $8.5\times 10^{-49}$ and $6.1\times 10^{-48}$ Hz$^{-1}$ over the sky and between  $1.2\times 10^{-48}$ and $1.2\times 10^{-47}$ Hz$^{-1}$ $\left[ 100 {\rm{Hz}}\over f\right]^3$  for a $\beta=-3$ spectrum.

\section{Continuous wave searches}

Fast rotating neutron stars are expected to emit a continuous \gw signal if they present a deviation from a perfectly axisymmetric shape, if their r-modes are excited, or if their rotation axis is not aligned with their symmetry axis. For a review of emission mechanisms and the detectability of the resulting signals see \cite{S2Fstat} and references therein. In all cases the expected signal at any given time is orders of magnitude smaller than any of the short-lived signals that have been described in sections \ref{cbc} and \ref{bursts}. However, since the signal is present for a very long time (to all practical purposes, in most cases, one may consider it there {\em{all}} the time), one can increase the SNR by integrating for a suitably long time. In its simplest form this is the problem of looking for a sinusoidal signal in Gaussian noise: the square modulus of the Fourier Transform of the data exhibits a peak at the signal frequency whose height with respect to the noise increases linearly with the time baseline of the data.

The waveform expected for this type of signal is well modeled. The signal at the detector depends on the position of the source, its rotational phase evolution as well as on the source polarization and orientation parameters, all or some of which may be unknown. In the case of many known pulsars the position and frequency evolution of the \gw signal is known from radio observations. Based on this, the \gw search is relatively straightforward \cite{S1,knownS2,knownS3S4}. 
No gravitational wave signal has been detected while searching for continuous \gws from known radio pulsars. This is not unexpected because for most systems the indirect upper limit on the amplitude of \gws that one may infer from the measured spin-down rate of the systems is more constraining that the limit determined by the \gw observations. However in one case \gw observations are actually beating the electromagnetic spin-down limit and starting to probe new ground. This is the case of the Crab pulsar. With several months of data at the sensitivity of the 5th science run it is expected that LIGO observations will beat the spin-down upper limit by a factor of a few. On other pulsars, albeit not beating the spin-down upper limits, the LIGO results are expected to reach values as low as a few $10^{-26}$ in $h_0$ and several $10^{-8}$ in ellipticity $\epsilon$. These results show that at the current sensitivity, it is possible that LIGO could detect a continuous \gw signal, coming from an unusually nearby object, unknown electromagnetically and rotating close to $\sim 75$ Hz.

The most promising searches look for previously unknown objects, and are often refered to as blind searches (\cite{S2Fstat,S2Hough,S4PHS}). Such searches cannot be carried out employing matched filtering on long data stretches due to the very large parameter space and the resulting unmanageable computational expense \cite{houghMethod}. One has then to resort to hierarchical procedures of various kinds, differently optimized depending on the type of signal. Searches that begin with a short time baseline coherent integration are robust with respect to frequency glitches shorter than the inverse of the coherent integration time and are relatively fast, requiring dedicated large  computer clusters (with of order $10^3$ CPU cores) for a few weeks (\cite{S2Hough,S4PHS}). The most sensitive searches begin with a long baseline coherent search, of order a few days, and require the signal to maintain phase coherence over such timescales. These deep searches need an enormous amount of computational power and in fact are carried out by Einstein@Home, a public distributed computing project that uses compute cycles donated by the general public. Einstein@Home is the second largest public compute project in the world and delivers an average 100Tflops of compute power continuously \cite{E@h}.

In the absence of a detection, upper limits are placed on the intrinsic amplitude of the \gw signal at the detector, $h_0$ (see Eq. 3.3 of \cite{S1}). Note that $h_0$ is {\em{not}} the amplitude of the \gw signal that excites the detector. The latter also depends on the relative position and orientation of the detector and the source and the polarization of the source. However it is convenient to express the upper limit of the searches in terms of $h_0$ because this quantity is easily connected to important physical properties of the source. If the \gw emission is due to a deviation from an axisymmetric shape, then
\begin{equation}
h_0={{4\pi^2 G\over c^4}{I \nu^2\over d}\epsilon}
\label{eq:h0}
\end{equation}
where $d$ is the distance to the source, $I$ is its principal moment of inertia about the rotation axis, $\nu$ the \gw frequency, $c$ the speed of light and $G$ Newton's constant. 

In the case of signals from known pulsars, one can easily translate $h_0$ in the ellipticity $\epsilon$ at a fiducial moment of inertia or in $\epsilon I$, the quadrupole moment. In the case of null results from blind searches the upper limits on $h_0$ are given per frequency interval and refer to populations of signals with certain priors on the unknown position, inclination, polarization and phase angles (typically uniform). Under the assumption that all the observed spindown power is emitted in \gws the following relation holds:
\begin{equation}
\epsilon= {7.6\times 10^5 ~ {\dot{\nu}}^{1/2}\over \nu^{5/2}}.
\label{eq:epsilonoffandfdot}
\end{equation}

\begin{figure}[!htbp]
\begin{center}
\includegraphics[width=0.95\textwidth]{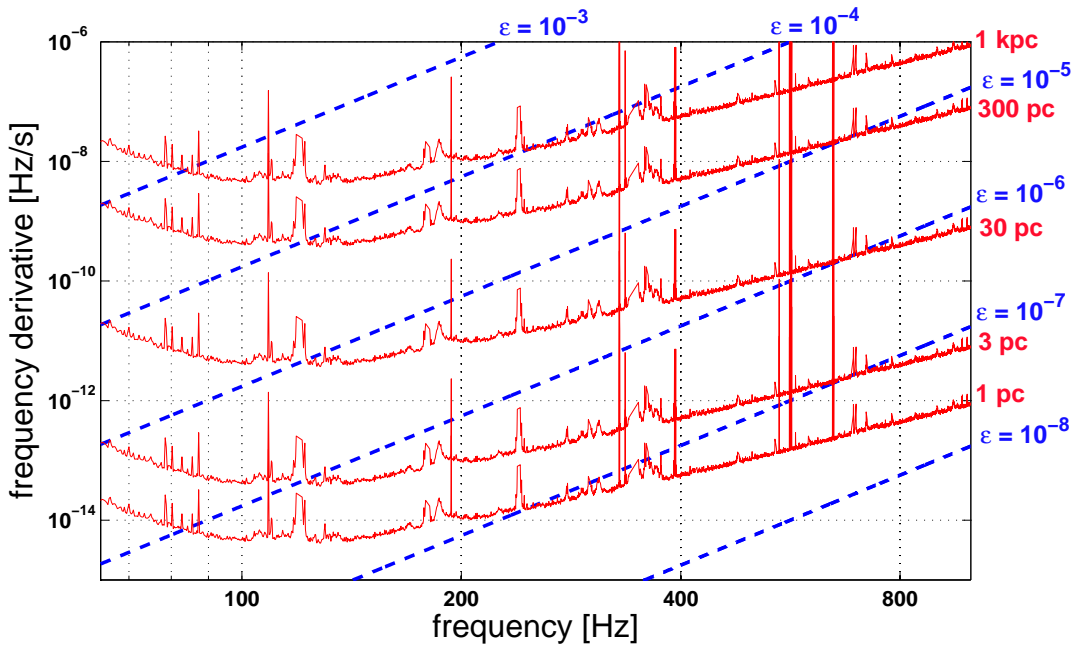}
\caption{Solid curves: Frequency and frequency-derivative values of a signal that would be detectable by the S4 stack-slide search described in \cite{S4PHS}. Dashed lines: lines of constant ellipticity.}
\label{fig:blindSearchReach}
\end{center}
\end{figure}

Eq. \ref{eq:epsilonoffandfdot} can be substituted in Eq.\ref{eq:h0} to derive an expression that depends only on $h_0,d,\nu$ and $\dot \nu$.
Using this expression the $h_0$ upper limits may be recast as contour plots in the frequency-first frequency derivative plane which represent a detectable \gw signal at a fixed distance. On the same plane one can overlay the $\epsilon = {\rm{constant}}$ curves and understand what ellipticity values the distance parametrized curves refer to. An example of this plot is Fig.{41} of \cite{S4PHS}. Fig.\ref{fig:blindSearchReach} shows a similar type of curve, deduced from the Hanford 4km interferometer S4 data stack-slide search upper limits of \cite{S4PHS}. In S5, the most sensitive Einstein@Home searches are expected to yield a sensitivity improvement in $h_0$ close to a factor of 10, resulting in a detectability range of $\sim$ 1 kpc at $150$ Hz with $\epsilon \sim 10^{-5}$.

\section{Conclusions}

Someone commented at the end of my talk in Sydney ``Don't you worry that when the only thing that you have is a hammer everything looks like a nail ?''.
I think that \gw research is nowdays well passed the stage where the hammer-nail analogy holds. In fact, even a brief review like this one shows the variety of \gw signals and emission scenarios that the last generation of \gw detectors can pursue. I look forward to the time when gravitational wave observations become routine and to being surprised by The New and Unexpected. However I also believe that the large body of indirect evidence regarding the nature and the emission mechanisms of gravitational waves, should reassure us that many of the signals that we look for are actually very close to what is out there.

Although \gws have not yet been detected, the most recent upper limits are starting to contribute new astronomical information. The S5 upper limit on the intrinsic \gw amplitude $h_0$ from the Crab pulsar will constrain the \gw emission below what any other observation can do, and place an informative constraint on the \gw contribution to the energy budget of this object. The S5 upper limit on an isotropic background of \gws will constrain $\Omega_{GW}$ below the big bang nucleosynthesis limit, which is a landmark result for this type of searches. Inspiral searches around the time of the GRB 070201 event have excluded with very high confidence the association of this GRB with a neutron star-neutron star or neutron star-black hole inspiral in Andromeda. 

Various types of \gw signals could be detected now that would not challenge the basic understanding that we have of astronomy, astrophysics or cosmology. In two years the enhanced detectors will have increased the volume of Universe that we can see by a factor of $\sim$ 8. In six years this volume will have increased by a factor of 1000. At that point cherished beliefs will have to be questioned if the data do not reveal any \gw signal.

\ack{
I am grateful to many collegues in the LIGO Scientific Collaborations for valuable discussions. I would especially like to thank P. Shawhan and A. Dietz for their reading of this manuscript and their useful comments.  
}

\end{document}